\documentclass[12pt]{iopart}

\usepackage{cite}

\usepackage{graphicx}
\usepackage[german, english]{babel}
\usepackage{color}

\begin{document}

\title{Superfluid quenching of the moment of inertia in a strongly interacting Fermi gas}


\author{S Riedl$^{1,2}$\footnote{Present address: Max-Planck-Institut f\"ur Quantenoptik, Garching, Germany.},
    E R S\'{a}nchez Guajardo$^{1,2}$,
    C Kohstall$^{1,2}$,
    J Hecker Denschlag$^{1}\footnote{Present address: Institut f\"ur Quantenmaterie, Universit\"at Ulm, Germany.}$,
    and R Grimm$^{1,2}$}

\address{$^1$ Institut f\"ur Experimentalphysik und Zentrum f\"ur Quantenphysik, Universit\"at Innsbruck, 6020 Innsbruck, Austria}
\address{$^2$ Institut f\"ur Quantenoptik und Quanteninformation, \"Osterreichische Akademie der Wissenschaften, 6020 Innsbruck, Austria}

\begin{abstract}
We report on the observation of a quenched moment of inertia as resulting from superfluidity in a strongly interacting Fermi gas. Our method is based on setting the hydrodynamic gas in slow rotation and determining its angular momentum by detecting the precession of a radial quadrupole excitation. The measurements distinguish between the superfluid and collisional origins of hydrodynamic behavior, and show the phase transition.
\end{abstract}

\maketitle


\section{Introduction}

Superfluidity is a striking property of quantum fluids at very low temperatures. For bosonic systems, important examples are liquids and clusters of $^4$He and atomic Bose-Einstein condensates. In fermionic systems, superfluidity is a more intricate phenomenon as it requires pairing of particles. Fermionic superfluidity is known to occur in atomic nuclei and $^3$He liquids and it is also at the heart of superconductivity, thus being of great technological importance. Recent advances with ultracold Fermi gases have opened up unprecedented possibilities to study the properties of strongly interacting fermionic superfluids \cite{Giorgini2008tou,Inguscio2006ufg}.
Early experiments on ultracold Fermi gases with resonant interparticle interactions compiled increasing evidence for superfluidity \cite{Ohara2002ooa,Regal2004oor,Kinast2004efs,Bartenstein2004ceo,Chin2004oop,Kinast2005hco} until the phenomenon was firmly established by the observation of vortex lattices \cite{Zwierlein2005vas}.


 Here we report on the manifestation of superfluidity in a quenched moment of inertia (MOI) in a strongly interacting Fermi gas that undergoes slow rotation. The basic idea of a quenched MOI as a signature of superfluidity dates back to more than 50 years ago in nuclear physics, where MOIs below the classical, rigid-body value were attributed to superfluidity \cite{Ring1980tmn}. The quenching of the MOI was also shown in liquid $^4$He \cite{Hess1967moa} and has, more recently, served for the discovery of a possible supersolid phase \cite{Kim2004oos}. Here we introduce the observation of the quenched MOI as a new method to study superfluidity in ultracold Fermi gases.

\section{Basic idea of the measurement}

\begin{figure}
\begin{center}
\includegraphics[width=.4\columnwidth,clip]{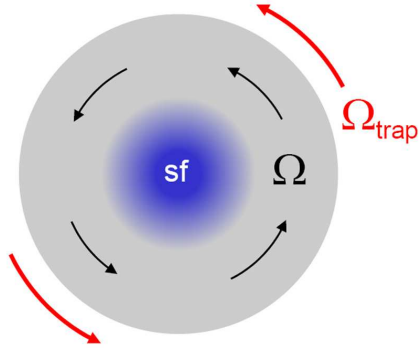}
\caption{Schematic illustration of a strongly interacting Fermi gas
in a slowly rotating trap. The normal part rotates with a frequency $\Omega$, which in an equilibrium state approaches the rotation frequency $\Omega_{\rm trap}$ that is imposed by the trap. The superfluid core (sf) does not carry angular momentum and therefore does not contribute to the MOI.}\label{idea}
\end{center}
\end{figure}

The basic situation that underlies our experiments is illustrated in Fig.~\ref{idea}. At a finite temperature below the critical temperature $T_c$, the harmonically trapped cloud consists of a superfluid core centered in a collisionally hydrodynamic cloud. We assume that the trapping potential is close to cylindrical symmetry, but with a slight, controllable deformation that rotates around the corresponding axis with an angular velocity $\Omega_{\rm trap}$. The nonsuperfluid part of the cloud is then subject to friction with the trap and follows its rotation with an angular velocity $\Omega$ \cite{rigid}, which in a steady state ideally reaches $\Omega = \Omega_{\rm trap}$. The corresponding angular momentum can be expressed as $L = \Theta \Omega$, where $\Theta$ denotes the MOI. The superfluid core cannot carry angular
momentum, assuming that vortex nucleation is avoided, and therefore does not contribute to the MOI of the system. Thus $\Theta$ represents the MOI of the whole system.

The case of a rotating system in a steady state, where the normal part carries the maximum possible angular momentum, allows us to distinguish the superfluid quenching of the MOI from a non-equilibrium quenching effect as studied in Ref.~\cite{Clancy2007oon}. There the authors investigated the hydrodynamic expansion of a gas with a known angular momentum. This situation, where the velocity fields of the normal and superfluid components are not in a steady state, can also be discussed in terms of a MOI below the rigid-body value. In contrast to the phenomenon investigated in our present work, the effect of Ref.~\cite{Clancy2007oon} is related to irrotational flow and can occur for both the superfluid and the collisionally hydrodynamic normal phase.


Our measurements rely on the possibility to determine the total angular momentum $L$ of a rotating hydrodynamic cloud by detecting the precession of a radial quadrupole excitation. This method is well established and has been extensively used in the context of atomic Bose-Einstein condensates \cite{Chevy2000mot,Haljan2001uos,Leanhardt2002ivi}. We have recently applied it to a rotating, strongly interacting Fermi gas to investigate the slow decay of angular momentum \cite{Riedl2009loa}.
The method works under the general condition that the gas behaves hydrodynamically. Then the precession frequency can be written as $\Omega_{\rm prec} = L / (2 \Theta_{\rm rig})$ \cite{Zambelli1998qva}, where $\Theta_{\rm rig}$ corresponds to a moment of inertia as calculated from the density distribution under the assumption that the whole cloud, including the superfluid part, would perform a rigid rotation.
Substituting $\Theta \Omega$ for $L$, we obtain $\Omega_{\rm prec} = {\Theta}/({2 \Theta_{\rm rig}}) \,\, \Omega$, with $\Theta/\Theta_{\rm rig} = 1$ for the full MOI in a normal system, and $\Theta/\Theta_{\rm rig} < 1$ for a MOI that is quenched because of the superfluid core.


\section{Experimental setup and procedures}
\label{sec:procs}

The starting point of our experiments is an optically trapped, strongly interacting Fermi gas consisting of an equal mixture of $^6$Li atoms in the
lowest two atomic states \cite{Jochim2003bec,Altmeyer2007doa}. The broad 834-G Feshbach resonance \cite{Inguscio2006ufg} allows us
to control the $s$-wave interaction. If not otherwise stated, the
measurements presented here refer to the resonance center. Here a unitarity-limited Fermi gas \cite{Giorgini2008tou,Inguscio2006ufg} is realized, which is known to exhibit deep hydrodynamic behavior even well above the critical temperature for superfluidity, see e.g.\ \cite{Wright2007ftc}. The cigar-shaped quantum gas is confined in a far red-detuned, single-beam optical dipole trap with additional axial magnetic confinement. The trap can be well approximated by a harmonic potential with radial oscillation frequencies $\omega_x=\omega_y\approx2\pi\times680$\,Hz and an axial frequency of $\omega_z=2\pi\times24$\,Hz. The Fermi energy of the noninteracting gas is given by $E_F=\hbar(3N\omega_x\omega_y\omega_z)^{1/3}$, where $N=6\times10^5$ is the total atom number. The Fermi temperature is
$T_F=E_F/k=1.3\,\mu$K, with $k$ denoting the Boltzmann constant.


Our scheme to study the rotational properties is described in detail in Ref.~\cite{Riedl2009loa}. It is based on a rotating elliptical deformation of the trap, characterized by a small ellipticity parameter \cite{Riedl2009loa} $\epsilon'=0.1$. In contrast to our previous work, we use a lower rotation frequency of $\Omega_{\rm trap} = 2\pi \times 200\,$Hz $\approx 0.3\,\omega_x$. This low value allows us to avoid a resonant quadrupole mode excitation, which is known as an efficient mechanism for vortex nucleation \cite{Madison2001sso,Hodby2002vni}. To excite the quadrupole mode \cite{Altmeyer2007doa} we switch on an elliptic trap deformation for 50\,$\mu$s. We detect the resulting oscillation by taking absorption images of the cloud after a variable hold time in the trap and a short free expansion time after release from the trap. More details on this excitation and detection scheme are given in Ref.~\cite{Riedl2009loa}.

At this point it is important to discuss the consequences of residual trap imperfections, still present when we attempt to realize a cylindrically symmetric optical potential. As we showed in previous work \cite{Riedl2009loa}, we can control the ellipticity down to a level of $\sim$1\%. Moreover, deviations from perfect cylindrical symmetry may occur because of other residual effects, such as corrugations of the optical trapping potential. As a consequence, a certain rotational damping is unavoidable, but damping times can reach typically one second \cite{Riedl2009loa}. This has two main effects for our observations. First, our measurements yield precession frequencies slightly below $\Omega_{\rm prec}$. This is because of a delay time of $20$\,ms between turning off the rotating trap ellipticity and applying the quadrupole mode excitation. It is introduced to make sure that any possible collective excitation resulting from the rotating trap has damped out when the mode precession is measured. Because of rotational damping during this delay time, the measured precession frequencies $\Omega'_{\rm prec}$ are somewhat below $\Omega_{\rm prec}$. To compensate for this effect, we directly measure the reduction of $\Omega_{\rm prec}$ that occurs during a 20\,ms hold time to determine the corresponding damping parameter $\kappa = \Omega'_{\rm prec}/\Omega_{\rm prec}$ for each set of measurements, finding day-to-day variations with typical values between $0.85$ and $0.9$. The second effect is induced by friction with static (nonrotating) trap imperfections when the rotating ellipticity is applied. This leads to equilibrium values for $\Omega$ typically a few percent below $\Omega_{\rm trap}$, depending on the ratio between the time constants for spin up and damping \cite{Odelin2000sua}. For this second effect there is no straightforward compensation, and it needs to be explicitly discussed when interpreting the experimental results.


\begin{figure}
  \begin{center}
  \includegraphics[width=.5\columnwidth,clip]{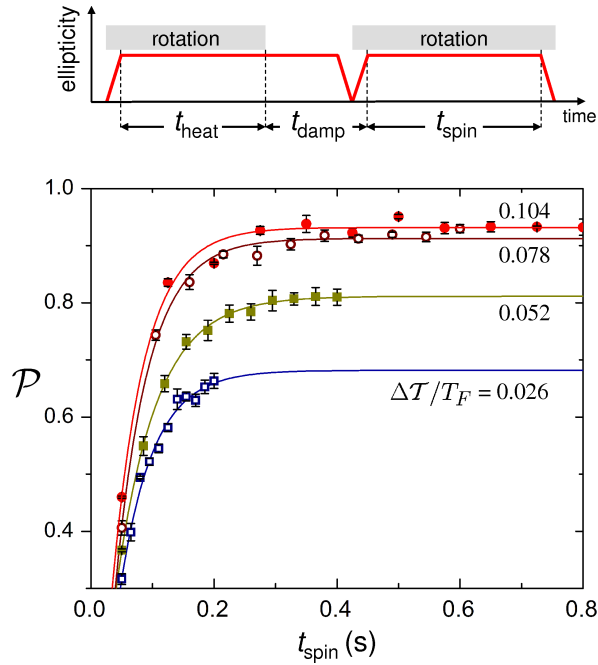}
  \caption{Precession parameter $\mathcal{P}$ versus spin-up time $t_{\rm spin}$ for various values of the final temperature, as characterized by the heating parameter $\Delta \mathcal{T}$ (see text). The quenching of the MOI shows up in the temperature-dependent saturation behavior. The applied timing sequence to facilitate measurements at constant temperature is illustrated above the graph. For these sets of measurements $\kappa = 0.85$.}\label{quench}
\end{center}
\end{figure}


Thermometry is performed after the whole experimental sequence. We damp out the rotation by stopping the trap rotation and keeping the ellipticity \cite{norotheat}. We convert the gas into a weakly interacting one by a slow magnetic field ramp to $1132$\,G, and we finally measure the temperature $\mathcal{T}$ \cite{Riedl2009loa}. Note that the isentropic conversion tends to decrease the temperature such that $\mathcal{T}$ is always somewhat below the temperature $T$ at unitarity \cite{Chen2005toi}.
The relative statistical uncertainty of the temperature measurement is about
$5\%$ in the relevant temperature range.


\section{Experimental results}


To discuss our experimental results we introduce a dimensionless precession parameter $\mathcal{P}$ by normalizing our observable $\Omega_{\rm prec}$ to its maximum possible value of $\Omega_{\rm trap} / 2$,
\begin{equation}
 \mathcal{P} = 2 \,\, \frac{\Omega_{\rm prec}}{\Omega_{\rm trap}} = \frac{\Theta}{\Theta_{\rm rig}} \times \frac{\Omega}{\Omega_{\rm trap}} \,.
\end{equation}
The maximum possible value of $\mathcal{P} = 1$ corresponds to a fully rotating, classically hydrodynamic cloud.
Values $\mathcal{P} < 1$ show the presence of at least one of the two effects, namely the incomplete rotation of the normal part ($\Omega / \Omega_{\rm trap} <1$) or the superfluid quenching of the MOI ($\Theta / \Theta_{\rm rig} <1$). It is crucial for the interpretation of our experimental results to distinguish between these two effects. Our basic idea to achieve this relies on the fact that $\Theta / \Theta_{\rm rig}$ represents a temperature-dependent {\it equilibrium} property, whereas $\Omega / \Omega_{\rm trap}$ depends on the dynamics of the spin-up before the system has reached an equilibrium.
Experimentally, however, measurements of equilibrium properties at a fixed temperature are not straightforward because of the presence of residual heating leading to a slow, steady temperature increase. In the rotating trap we always observe some heating, which under all our experimental conditions can be well described by a constant rate $\alpha = 170$\,nK/s = $0.13\, T_F$/s \cite{measureheating}.

\subsection{Equilibrium state of rotation}
\label{sec:equilib}

To identify the conditions under which our cloud reaches its equilibrium state of rotation, we have developed a special procedure based on the timing scheme illustrated on top of Fig.\,\ref{quench}. Our procedure takes advantage of the constant heating rate $\alpha$ to control the final temperature of the gas when $\mathcal{P}$ is measured. We apply the trap rotation in two separate stages of duration $t_{\rm heat}$ and $t_{\rm spin}$. In an intermediate time interval of $t_{\rm damp} = 200$\,ms \cite{tdamp} we damp out the rotation that is induced by the first stage. The angular momentum disappears, but the heating effect remains \cite{norotheat}. The second stage spins up the cloud again and induces further heating. When $t_{\rm tot} = t_{\rm heat}+ t_{\rm spin}$ is kept constant, we find that the total heating by the two rotation stages is $\Delta \mathcal{T} = \alpha \, t_{\rm tot}$. As only the second stage leads to a final angular momentum, the equilibrium state reached at a constant temperature can be identified when $\mathcal{P}(t_{\rm spin})$ reaches a constant value for increasing $t_{\rm spin}$ and fixed $t_{\rm tot}$. The temperature can be controlled by a variation of the parameter $t_{\rm tot}$ and is obtained as $\mathcal{T} = \mathcal{T}_0 + \Delta \mathcal{T}$. The temperature offset $\mathcal{T}_0$ is set by the initial cooling and some unavoidable heating during the experimental sequence without trap rotation. Under our conditions $\mathcal{T}_0 \approx 0.11 \, T_F$.

Our experimental results for $\mathcal{P}(t_{\rm spin})$ are shown in Fig.\,\ref{quench} for four different values of the heating parameter $\Delta \mathcal{T}/T_F$ in a range between $0.026$ to $0.104$, which corresponds to a range of $\mathcal{T}$ between about $0.14$ and $0.21\,T_F$. All four curves show qualitatively the same behavior. Within a few 100\,ms, $\mathcal{P}$ rises before reaching a final equilibrium value. This time-dependent increase of $\mathcal{P}$ is related to the spin-up dynamics \cite{spinupcomplicated}. We find that the observed increase and saturation of $\mathcal{P}(t_{\rm spin})$ can be well fit by simple exponential curves (solid lines), and we use these fits to extract the different equilibrium values $\mathcal{P}_{\rm eq}$.

The equilibrium values $\mathcal{P}_{\rm eq}$ exhibit an interesting temperature dependence. The lower three values show a pronounced increase with temperature, $\mathcal{P}_{\rm eq} = 0.68$,  $0.81$, and $0.91$ for $\Delta\mathcal{T}/T_F = 0.026$, $0.052$, and $0.078$, respectively. We interpret this increase as a consequence of the decreasing superfluid core and thus the decreasing MOI quenching effect. For our highest temperature ($\Delta\mathcal{T}/T_F = 0.104$) we only observe a marginal further increase to $\mathcal{P}_{\rm eq} = 0.93$. This indicates that the superfluid core is very small or absent leading to a disappearance of the quenching effect. The fact that the maximum $\mathcal{P}_{\rm eq}$ stays a few percent below $1$ can be explained by trap imperfections as discussed in Sec.\ \ref{sec:procs}.


Let us comment on the possible influence of vortices \cite{Zwierlein2005vas}. We cannot exclude their presence \cite{vortices}, as their nucleation can proceed not only via a resonant quadrupole mode excitation \cite{Madison2001sso,Hodby2002vni}, but also via a coupling to the thermal cloud \cite{Haljan2001dbe}. Vortices would result in additional angular momentum in the rotating cloud and its collective behavior would be closer to the normal case. This would tend to increase $\mathcal{P}$ at lower temperatures, counteracting the behavior that we observe.


\subsection{Superfluid phase transition}

In a second set of experiments, we study the superfluid phase transition in a way which is experimentally simpler, but which requires information on the equilibrium state as obtained from the measurements presented before.
The trap rotation is applied continuously, and we observe the increase of $\mathcal{P}$ with the rotation time $t_{\rm rot}$. All other parameters and procedures are essentially the same as in the measurements before. Here the temperature is not constant, but rises according to $\mathcal{T} = \mathcal{T}_0 + \alpha t_{\rm rot}$, where the heating rate $\alpha = 170\,$nK/s is the same as before and $\mathcal{T}_0 = 0.085 T_F$ is somewhat lower because of the less complex timing sequence.

\begin{figure}
  \begin{center}
  \includegraphics[width=.5\columnwidth,clip]{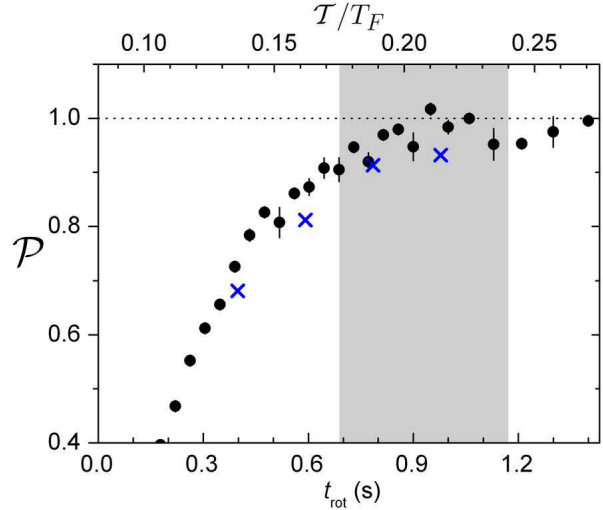}\\
  \caption{The precession parameter $\mathcal{P}$ as a
  function of the rotation time $t_{\rm rot}$ (filled symbols); the upper scale shows the corresponding temperature $\mathcal{T}$. For comparison, the crosses show the equilibrium values $\mathcal{P_{\rm eq}}$ as obtained from Fig.\ \ref{quench}. The shaded region indicates the range in which we expect the superfluid phase transition according to previous experiments \cite{Regal2004oor,Luo2007mot,Inada2008cta,Luo2009tmi,Nascimbene2010ett,Horikoshi2010mou}. For this set of measurements $\kappa = 0.90$.}\label{spinup}
\end{center}
\end{figure}

Figure \ref{spinup} shows how $\mathcal{P}$ increases with the rotation time $t_{\rm rot}$ (filled symbols); the upper scale shows the corresponding temperature $\mathcal{T}$. The observed increase of $\mathcal{P}$ generally results from both factors in Eq.(\ref{idea}), corresponding to the rising $\Omega/\Omega_{\rm trap}$ (spin-up dynamics) and the rising $\Theta/\Theta_{\rm rig}$ (decrease of the superfluid MOI quenching). Figure \ref{spinup} also shows the values $\mathcal{P_{\rm eq}}$ as determined from Fig.\ \ref{quench} (crosses), for which we know that the spin-up of the normal component has established an equilibrium with $\Omega/\Omega_{\rm trap}$ being close to one. The comparison shows that already for $t_{\rm rot} =0.4$\,s the data set obtained with the simpler procedure follows essentially the same behavior. The small quantitative difference that the crosses are slightly below the open symbols can be explained by a somewhat stronger influence of trap imperfections in the earlier measurements of Sec.\ \ref{sec:equilib} \cite{imperfect} or by the uncertainty in the initial temperature $\mathcal{T}_0$. For $t_{\rm rot} \ge 0.4$\,s, we can assume that the system is in an equilibrium state, which follows the slowly increasing temperature, and we can fully attribute the further increase of $\mathcal{P}$ to the quenching of the MOI.

The superfluid phase transition corresponds to the point where the precession parameter $\mathcal{P}$ reaches its saturation value. This is observed for a time $t_{\rm rot} \approx 0.95$\,s, when $\mathcal{T}/T_F \approx 0.21$. The conversion of this temperature parameter (measured in the weakly interacting regime after an isentropic change) to the actual temperature in the unitarity-limit regime \cite{Luo2009tmi} yields a value for the critical temperature $T_c$ of about $0.2\,T_F$. This result is consistent with previous experimental results \cite{Regal2004oor,Luo2007mot,Inada2008cta,Luo2009tmi,Horikoshi2010mou,Nascimbene2010ett}, the range of which is indicated by the shaded region in Fig.\ \ref{spinup}. The result is also consistent with theoretical predictions \cite{Giorgini2008tou,Haussmann2008toa}.

For a more precise extraction of $T_c$ from experimental MOI quenching data, a theoretical model would be required that describes the saturation behavior of $\Theta/\Theta_{\rm rig}$ as $T_c$ is approached. Theoretical predictions are available for the BEC limit \cite{Stringari1996moi} and the BCS limit \cite{Farine2000moi,Urban2003sro,Urban2005tfm}. In the unitarity limit it should, in principle, be possible to extract the MOI from spatial profiles of the normal and the superfluid fraction \cite{Perali2004bbc,Stajic2005dpo}. Clearly, more work is necessary to quantitatively understand the quenching effect in the strongly interacting regime.


\section{Conclusion}

We have demonstrated the quenching of the moment of inertia that occurs in a slowly rotating, strongly interacting Fermi gas as a consequence of superfluidity. This effect provides us with a novel probe for the system as, in contrast to other common methods such as expansion measurements and studies of collective modes, it allows us to distinguish between the two possible origins of hydrodynamic behavior, namely collisions in a normal phase and superfluidity.

\section*{Acknowledgments}

We thank L.\ Sidorenkov and M.\ K.\ Tey for discussions. We acknowledge support by the Austrian Science Fund (FWF) within SFB 15 (project part 21) and SFB 40 (project part 4).

\section*{References}


\begin{thebibliography}{10}
\bibitem{rigid}
We assume that the normal cloud performs a rigid rotation with an angular
  velocity $\Omega$. This can be justified by the internal friction in the
  non-superfluid component along with the fact that the rotating trap
  deformation is applied to all regions of the cloud simultaneously.
\bibitem{norotheat}
 The temperature increase resulting from conversion of rotational energy into
  heat is neglibly small.
\bibitem{measureheating}
To determine the temparature increase in the rotating trap we
  measure $\mathcal{T}$ after variable rotation times. We find that, in the
  relevant temperature range, the behavior of $\mathcal{T}$ is well
  approximated by a linear increase with time. This justifies the description
  in terms of a constant heating rate.
\bibitem{tdamp}
 The ellipticity is kept at its full level while the rotation is turned off. To
  speed up the damping we increase the magnetic field to 920G.
\bibitem{spinupcomplicated}
The curves do not show the spin-up process directly, as our measurement
  procedure fixes the temperature at the time of the measurement of
  $\mathcal{P}$.
\bibitem{imperfect}
This explanation is supported by the fact that we measured a slower decay of
  angular momentum for the later experiments of Fig.\ \ref{spinup}
  ($\kappa=0.90$) than we did for the earlier measurements of Fig.\
  \ref{quench} ($\kappa=0.85$). Between the two sets of measurements the
  optical setup of the trapping beam was readjusted, leading to reduced
  imperfections in the later experiments.
\end{thebibliography}
\providecommand{\newblock}{}

\end{document}